\documentclass[aps,pra,twocolumn,superscriptaddress,longbibliography]{revtex4-1}
\usepackage{lineno,hyperref,amsmath}
\usepackage{booktabs}
\usepackage{graphicx}
\usepackage{subfigure}
\usepackage{float}
\usepackage{amsmath}
\usepackage{amssymb}
\usepackage{appendix}
\usepackage{tikz}
\usepackage{bm}

\hypersetup{hidelinks,
colorlinks=true,
allcolors=black,
pdfstartview=Fit,
breaklinks=true}

\definecolor{lime}{HTML}{A6CE39}
\DeclareRobustCommand{\orcidicon}{
\begin{tikzpicture}
\draw[lime, fill=lime] (0,0)
circle[radius=0.16]
node[white]{{\fontfamily{qag}\selectfont \tiny \.{I}D}};
\end{tikzpicture}
\hspace{-2mm}
}
\foreach \x in {A, ..., Z}{%
\expandafter\xdef\csname orcid\x\endcsname{\noexpand\href{https://orcid.org/\csname orcidauthor\x\endcsname}{\noexpand\orcidicon}}
}

\DeclareGraphicsExtensions{.png,.jpg,.eps}

\begin{document}
\author{Wei Luo}
\affiliation{National Laboratory of Solid State Microstructures, School of Physics,
Jiangsu Physical Science Research Center, and Collaborative Innovation Center of Advanced Microstructures,
Nanjing University, Nanjing 210093, China}
\affiliation{School of Science, Jiangxi University of Science and Technology, Ganzhou 341000, China}

\author{Yating Zhang}
\affiliation{National Laboratory of Solid State Microstructures, School of Physics,
Jiangsu Physical Science Research Center, and Collaborative Innovation Center of Advanced Microstructures,
Nanjing University, Nanjing 210093, China}

\author{Wei Chen \hspace{-1.5mm}\orcidA{}}
\email{Corresponding author: pchenweis@gmail.com}
\affiliation{National Laboratory of Solid State Microstructures, School of Physics,
Jiangsu Physical Science Research Center, and Collaborative Innovation Center of Advanced Microstructures,
Nanjing University, Nanjing 210093, China}

\title{Manipulation of helical states revealed by crossed Andreev reflection}
\begin{abstract}
The edge states of a quantum spin Hall insulator exhibit helical properties, which has generated
significant interest in the field of spintronics. Although it is predicted theoretically that
Rashba spin-orbit coupling can effectively regulate edge state spin, experimental characterization
of this spin control remains challenging. Here, we propose utilizing crossed Andreev reflection to
probe the efficiency of gate voltage control on edge state spin. We calculate the transport
properties of the quantum spin Hall isolators-superconductor heterojunction by using non-equilibrium
Green's function method. We find that the probabilities of crossed Andreev reflection, electron
transmission, and non-local conductance all include the relative spin rotation angle $\theta$
caused by the Rashba spin-orbit coupling in the helical edge states. Notably, when the incident
energy is close to the superconducting gap, the differential conductance $G_2\propto-\cos{\theta}$.
Thus, the influence of Rashba spin-orbit coupling on edge state spin can be quantitatively
characterized by crossed Andreev reflection, which provides a feasible scheme for experimental testing.
\end{abstract}

\date{\today}

\maketitle

\section{Introduction}
Quantum spin Hall insulator (QSHI) is a topologically nontrivial phase of electronic matter, which
has a bulk insulating gap with gapless edge states traversing this
gap~\cite{Hasan10rmp,Qi11rmp,Bernevig06sci,Koenig07sci,Roth09sci,Kane05prl}. These edge states are
protected from impurity scattering by the nontrivial bulk band topology and time reversal symmetry. Due
to the spin-momentum locking~\cite{Wu06prl,Qi08prb} in the edge states, QSHIs have garnered significant
interest in the field of spintronics~\cite{Maciejko10prb,Krueckl11prl,Dolcini11prb,Modak12prb}. Compared
to traditional electronics, spintronics offers several advantages, including higher information density,
lower power consumption, and the potential for novel electronic devices and technologies~\cite{Wolf01sci,Sarma01rmp}.
Consequently, spintronics holds extensive application prospects in areas such as information
storage~\cite{Parkin08sci}, quantum computing~\cite{Meier03prl}, and sensing devices~\cite{Melzer11nano}.

A fundamental aspect of spintronics is spin control, which entails the precise manipulation of spin
direction and its dynamic evolution through external fields for applications in information storage and
processing. In QSHIs, spin-momentum locking manifests as a consequence of spin-orbit coupling (SOC) in the edge
states, indicating that the electron spin can be influenced by external fields that affect the spatial
wave function. By imposing a perpendicular electric field $E_z$, the Rashba spin-orbit coupling can
be generated~\cite{Rothe10njp,Strom10prl,Vayrynen11prl}, which provides a convenient and efficient method
for manipulating edge state spin~\cite{Sternativo14prb,Chen16prl,Ronetti16prb,Ortiz16prb,Garcia22prb}. The
control of Rashba SOC through gate voltage is a central topic in semiconductor spintronics~\cite{Rashba60spss},
benefiting from relatively mature technology that facilitates its integration into QSHIs. Although theoretical
predictions suggest that Rashba SOC can effectively regulate edge state spin, experimental characterization
of this spin control remains challenging due to its subtlety in conventional transport measurements. Therefore,
it is crucial to develop a viable and quantifiable approach to assess the efficiency of Rashba SOC in
controlling edge state spin.

\begin{figure}[t!]
\begin{center}
\includegraphics[width=1\columnwidth]{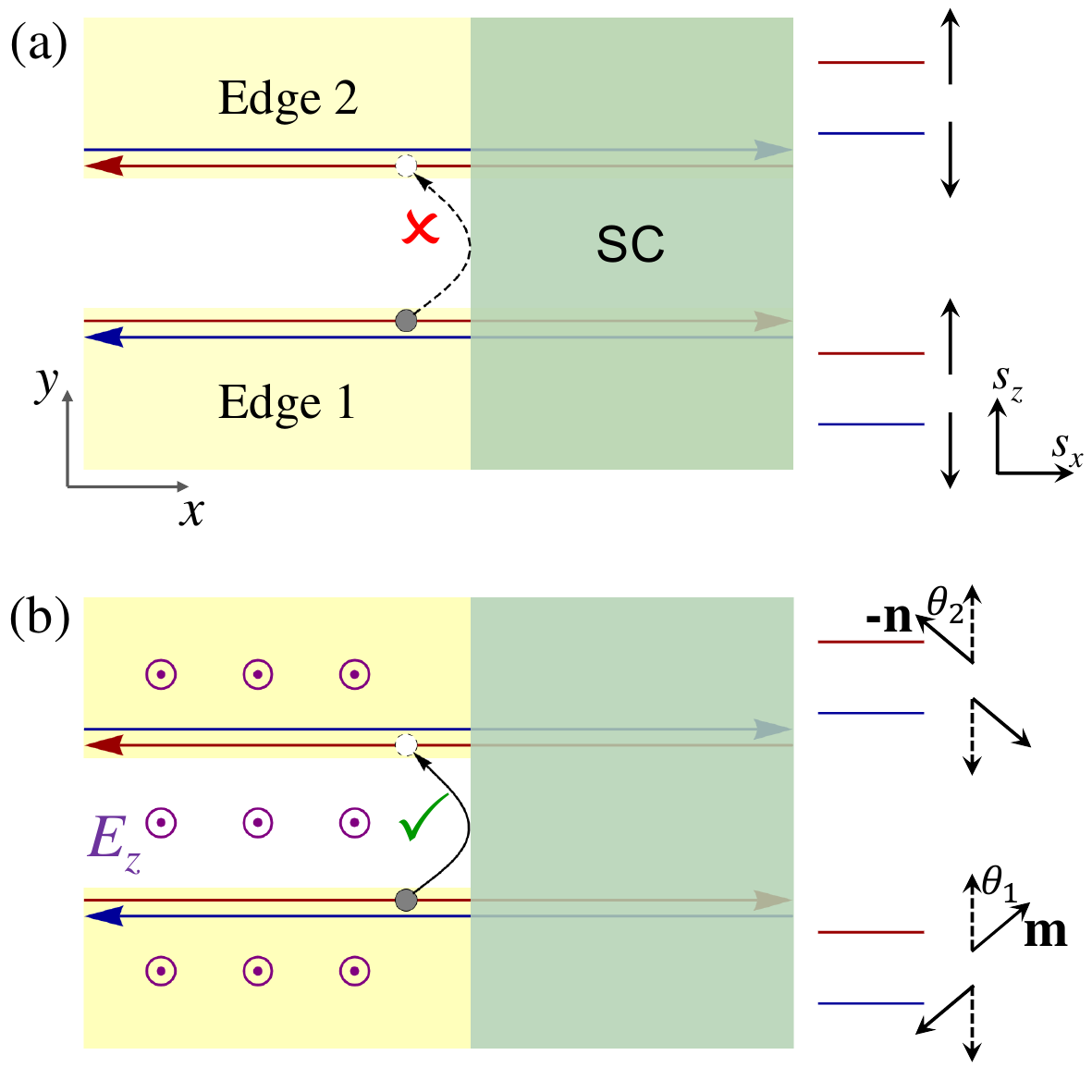}\\[0pt]
\caption{(Color online) Schematic illustration of the proposed structure. Two QSHIs in close proximity are in
contact with the $s$-wave SC. In (a), where there is no Rashba SOC, the opposite helicities of the adjacent
edge states prevent CAR. In (b), a perpendicular electric field $E_z$ induces Rashba SOC, causing the spin
rotation directions of adjacent edge states to be opposite. The spin of the incoming electrons at edge 1
rotates clockwise, while the spin of the outgoing state at edge 2 rotates counterclockwise. As a result, the
incoming state and the outgoing hole are no longer orthogonal, allowing CAR to occur.}
\label{Fig1}
\end{center}
\end{figure}

In this work, we propose utilizing crossed Andreev reflection (CAR)~\cite{Byers95prl,Deutscher00APL} to probe
the efficiency of gate voltage in regulating edge state spin. CAR has been extensively studied in
various electronic systems, as it may lead to important applications in quantum information
processing~\cite{lesovik01epjb,Recher01prb,Chen11prb,ChenW,Luow22prl} and exhibit interesting behaviors in various systems~\cite{Das24prb,Zeng22prb,Fuchs21prb}. The proposed setup is shown in Fig.\ \ref{Fig1},
two QSHIs close to each other are in contact with an $s$-wave superconductor (SC). Coherent reflection between
the QSHIs can occur if they are sufficiently close, e.g., with their distance being on the order of the
coherence length $\xi$ of the SC. $s$-wave SCs require the pairing of electrons with opposite spins.
Consequently, when an electron with spin $\mathbf{m}$ is incident, it attracts an electron with spin
$-\mathbf{m}$ into the SC, leaving behind a reflected hole with spin $\mathbf{m}$. In the absence of Rashba
SOC, adjacent edge states exhibit opposite helicities, as shown in Fig.\ \ref{Fig1}(a). Such spin configurations
prevent an incoming spin-up electron at edge 1 from pairing with a spin-down outgoing electron that is absent
at edge 2, thus inhibiting CAR. However, Rashba SOC exerts a uniform effect on the spins of both sets of edge
states, unlike helicity, which possesses opposite values for different edges. As a result, the spin polarization
directions of adjacent edge states rotate in opposite directions, as illustrated in Fig.\ \ref{Fig1}(b). Under
Rashba SOC, the spin of the incident electron at edge 1 rotates clockwise towards the $\mathbf{m}$ direction,
while the spin of the outgoing electron at edge 2 rotates counterclockwise towards the $-\mathbf{n}$ direction.
Since the $s$-wave SC pairs electrons with opposite spins, the CAR amplitude is expected to be proportional to $\langle-\mathbf{m}|-\mathbf{n}\rangle$,
with $|-\mathbf{m}\rangle$ ($|-\mathbf{n}\rangle$) representing the spin state polarized along the
$-\mathbf{m}$ ($-\mathbf{n}$) direction, which is related to the strength of the Rashba SOC. In this way, CAR
can be used to quantitatively characterize the effect of Rashba SOC on edge state spin, thereby providing a
viable approach for experimental verification.

The rest of the paper is organized as follows. In Sec.\ II we present the model and derive the transport
coefficients and current. The numerical results are discussed in Sec.\ III, and finally, we draw
our conclusions in Sev.\ IV.

\section{Model and calculations}
In the following, we calculate the transport properties of the QSHIs-SC heterojunction using non-equilibrium
Green's function method to obtain rigorous results. 
The setup in Fig.~\ref{Fig1} constitutes three parts, two closely-placed QSHIs, both covered by an $s$-wave
SC on their right halves. For typical QSHIs, the energy gaps of the bulk states are much larger than that of
the SC. Therefore, at the energy scale of the superconducting gap, only the helical edge states
are relevant. The Hamiltonian that effectively describes the device can be expressed as
\begin{equation}
  H=H_{\mathrm{E}}+H_{\mathrm{R}}+H_{\mathrm{S}}+H_{\mathrm{T}},
\end{equation}
where $H_{\mathrm{E}}$ describes the two adjacent but decoupled helical edge states, $H_{\mathrm{R}}$ accounts
for the Rashba SOC effect introduced in the edge states, $H_{\mathrm{S}}$ describes the $s$-wave
SC covering the right half of the QSHIs, and $H_{\mathrm{T}}$ represents the coupling between the helical edge
states and the SC.

Specifically, the one-dimensional Hamiltonian for the helical edge states is expressed as
\begin{equation}\label{Ham QSH}
H_{\mathrm{E}}=\sum_{\alpha p}s_{\alpha} vp(c_{\alpha p\uparrow}^{\dag}c_{\alpha p\uparrow}-
c_{\alpha p\downarrow}^{\dag}c_{\alpha p\downarrow}),
\end{equation}
where $c_{\alpha p\uparrow(\downarrow)}$ are the annihilation operators for electrons with spin
$\uparrow$ ($\downarrow$) and momentum $p$, and $v$ is the Fermi velocity. The edge index $\alpha=1,2$, and
$s_1=1$, $s_2=-1$ capture the spin helicities of the edge states. To
manipulate the spins of the edge states, a perpendicular electric field $E_z$ is applied to the QSHIs, which
induces Rashba SOC described by
\begin{equation}\label{Rashba}
H_{\mathrm{R}}=\lambda_{\alpha}p(c_{\alpha p\uparrow}^{\dag}
c_{\alpha p\downarrow}+c_{\alpha p\downarrow}^{\dag}c_{\alpha p\uparrow}),
\end{equation}
where the Rashba coefficient satisfies $\lambda_{\alpha}\propto eE_z$. The Rashba SOC is equivalent to a
nonuniform Zeeman field, which changes the spin directions of the helical edge states. Let us first examine
edge 1. By combining Eqs.\ (\ref{Ham QSH}) and (\ref{Rashba}), one obtains the eigenvalues
$\varepsilon_{1\pm}=\pm v_1p$, where $v_1=\sqrt{v^2+\lambda_{1}^{2}}$. The corresponding eigenstates
are given by
\begin{equation}
\begin{split}
|\mathbf{m}\rangle=&\cos{\frac{\theta_1}{2}}|\uparrow\rangle+\sin{\frac{\theta_1}{2}}|\downarrow\rangle,\\
|-\mathbf{m}\rangle=&-\sin{\frac{\theta_1}{2}}|\uparrow\rangle+\cos{\frac{\theta_1}{2}}
|\downarrow\rangle,\nonumber
\end{split}
\end{equation}
with $\theta_1=\arctan(\lambda_1/v)$. The electrons in states $|\mathbf{m}\rangle$
have spins parallel to the direction
$\mathbf{m}=\langle \mathbf{m}|\bm{\sigma}|\mathbf{m}\rangle=(\sin{\theta_1},0,\cos{\theta_1})$,
while the electrons in $|-\mathbf{m}\rangle$ have opposite spins. The Rashba SOC induces a clockwise
rotation of the spins of edge states by an angle $\theta_1$, as illustrated in Fig.\ \ref{Fig1}(b).
For edge 2, the eigenvalues are $\varepsilon_{2\pm}=\pm v_2p$, where $v_2=\sqrt{v^2+\lambda_{2}^{2}}$.
The corresponding eigenstates are
\begin{equation}
\begin{split}
|\mathbf{n}\rangle=\sin{\frac{\theta_2}{2}}|\uparrow\rangle+\cos{\frac{\theta_2}{2}}|\downarrow\rangle,\\
|-\mathbf{n}\rangle=\cos{\frac{\theta_2}{2}}|\uparrow\rangle-\sin{\frac{\theta_2}{2}}
|\downarrow \rangle,\nonumber
\end{split}
\end{equation}
with spins pointing to $\mathbf{n}=(\sin{\theta_2},0,-\cos{\theta_2})$ and
$-\mathbf{n}$, respectively, where $\theta_2=\arctan(\lambda_2/v)$. For edge 2, Rashba SOC causes a
counterclockwise rotation of the spin by an angle $\theta_2$, as shown Fig.\ \ref{Fig1}(b).

A two-dimensional SC is deposited on the right halves of the QSHIs. The distance $L$
between the two edge states is chosen to be smaller than the superconducting coherence length $\xi$,
ensuring that the variation of the quasiparticle wavefunction within the SC on this scale can be neglected~\cite{Deutscher00APL,Blonde82PRB}. Under this approximation, the distance $L$ between the
two QSHIs does not explicitly appear in the Hamiltonian of the SC, which is expressed as
\begin{equation}\label{Ham SC}
H_{\mathrm{S}}=\sum_{k\sigma}\varepsilon_kd_{k\sigma}^{\dag}d_{k\sigma}+\Delta\sum_{k}(
d_{k\uparrow}^{\dag}d_{-k\downarrow}^{\dag}+d_{-k\downarrow}d_{k\uparrow}),
\end{equation}
where $d_{k\sigma}$ are the annihilation operators for electrons with momentum $k$ and spin
$\sigma=\uparrow, \downarrow$, and $\Delta$ is the pair potential. The coupling between QSHIs
and SC is
\begin{equation}\label{cou}
H_{\mathrm{T}}=\sum_{\alpha pk\sigma}(T_{\alpha pk}c_{\alpha p\sigma}^{\dagger}d_{k\sigma}+
T_{k\alpha p}d_{k\sigma}^{\dagger}c_{\alpha k\sigma}),
\end{equation}
with $T_{\alpha kp}$ the tunneling strength, satisfying $T_{\alpha kp}=T_{p\alpha k}^{*}$.

It is convenient to calculate the transport properties using the spin eigenstate representations of edges
1 and 2. The corresponding bases are ($c_{1pm}$,$c_{1p\bar{m}}$) and ($c_{2pn}$,$c_{2p\bar{n}}$),
with subscripts $m$ ($\bar{m}$) representing states $|\mathbf{m}\rangle$ ($|-\mathbf{m}\rangle$), and
$n$ ($\bar{n}$) representing $|\mathbf{n}\rangle$ ($|-\mathbf{n}\rangle$). Additionally, the scattering
region includes the SC, so we will work in the Nambu space. The Hamiltonian of the helical edge states
is then expressed as
\begin{equation}\label{BDG QSH}
\begin{split}
\mathcal{H}_{\mathrm{E}}=&\sum_{p}(\tilde{c}_{1p}^{\dagger}h_{1p}\tilde{c}_{1p}+
\tilde{c}_{2p}^{\dagger}h_{2p}\tilde{c}_{2p}),\\
h_{\alpha p}=&\left(
\begin{array}{cccc}
v_{\alpha}p & 0 & 0 &0 \\
0 & -v_{\alpha}p & 0 & 0\\
0 & 0 & v_{\alpha}p & 0\\
0 & 0 & 0 & -v_{\alpha}p\\
\end{array}
\right),
\end{split}
\end{equation}
with Numba bases $\tilde{c}_{1p}=(c_{1pm},c_{1p\bar{m}},c_{1\bar{p}m}^{\dagger},
c_{1\bar{p}\bar{m}}^{\dagger})^{T}$ and
$\tilde{c}_{2p}=(c_{2pn},c_{2p\bar{n}},c_{2\bar{p}n}^{\dagger},c_{2\bar{p}\bar{n}}^{\dagger})^{T}$.
For simplicity, we express the superconducting Hamiltonian in the spin basis $|\pm\mathbf{m}\rangle$ as well
\begin{equation}\label{Ham SC2}
\begin{split}
\mathcal{H}_{\mathrm{S}}=&\sum_{k}\tilde{d}_{k}^{\dagger}h_{\mathrm{S}}\tilde{d}_{k},\\
h_{\mathrm{S}}=&\left(
\begin{array}{cccc}
\varepsilon_k & 0 & 0 &\Delta \\
0 & \varepsilon_k & -\Delta & 0\\
0 & -\Delta & -\varepsilon_{-k} & 0\\
\Delta & 0 & 0 & -\varepsilon_{-k}\\
\end{array}
\right),
\end{split}
\end{equation}
with $\tilde{d}_{k}=(d_{km},d_{k\bar{m}},d_{\bar{k}m}^{\dagger},d_{\bar{k}\bar{m}}^{\dagger})^{T}$.
The coupling Hamiltonian is given by
\begin{equation}\label{Coupling}
\mathcal{H}_{\mathrm{T}}=\sum_{pk}(\tilde{c}_{1p}^{\dagger}t_{1pk}\tilde{d}_{k}
+\tilde{c}_{2p}^{\dagger}t_{2pk}\tilde{d}_{k}+\mathrm{H.c.}),
\end{equation}
where the hopping matrices $t_{1pk}$ and $t_{2pk}$ are defined as
\begin{equation}
\begin{split}
\nonumber
t_{1pk}=\left(
\begin{array}{cccc}
T_{1pk} & 0 & 0 & 0 \\
0 & T_{1pk} & 0 & 0\\
0 & 0 & -T_{1\bar{p}\bar{k}}^{*} & 0\\
0 & 0 & 0 & -T_{1\bar{p}\bar{k}}^{*}\\
\end{array}
\right),
\end{split}
\end{equation}
and
\begin{equation}
\begin{split}
\nonumber
t_{2pk}=\left(
\begin{array}{cccc}
T_{2pk}^{nm} & T_{2pk}^{n\bar{m}} & 0 & 0 \\
T_{2pk}^{\bar{n}m} & T_{2pk}^{\bar{n}\bar{m}} & 0 & 0\\
0 & 0 & -T_{2\bar{p}\bar{k}}^{nm*} & -T_{2\bar{p}\bar{k}}^{n\bar{m}*}\\
0 & 0 & -T_{2\bar{p}\bar{k}}^{\bar{n}m*} & -T_{2\bar{p}\bar{k}}^{\bar{n}\bar{m}*}\\
\end{array}
\right),
\end{split}
\end{equation}
with $T_{2pk}^{nm}=\sin(\theta/2)T_{2pk}$, $T_{2pk}^{n\bar{m}}=\cos(\theta/2)T_{2pk}$, $T_{2pk}^{\bar{n}m}=\cos(\theta/2)T_{2pk}$, $T_{2pk}^{\bar{n}\bar{m}}=-\sin(\theta/2)T_{2pk}$,
and $\theta=\theta_1+\theta_2$.


We focus on the current in the edge 2 which is given by
\begin{equation}\label{cur1}
\begin{split}
I_2(t)=&-e\langle \dot{N}_2(t)\rangle=\frac{ie}{\hbar}\langle [N_2,H]\rangle\\
=&\frac{2e}{\hbar}\mathrm{Re}\Big\{\sum_{pk}\mathrm{Tr}[\tau_zt_{2pk}\mathbf{G}_{k2p}^{<}(t,t)]\Big\},
\end{split}
\end{equation}
where the Green's function $\mathbf{G}_{k2p}^{<}(t,t)]$ is defined as
$\mathbf{G}_{k2p}^{<}(t,t')=i\langle[\tilde{c}_{2p}^{\dagger}(t')]^T[\tilde{d}_k(t)]^{T}\rangle^{T}$.
Performing analytic continuation, the Green's function can be written as
\begin{equation}\label{analytic continuation}
\begin{split}
\mathbf{G}_{k2p}^{<}(t,t')=&\sum_{k'}\int dt_1\Big[\mathbf{G}_{kk'}^{R}(t,t_1)t_{2pk'}^{\dagger}(t_1)
\mathbf{g}_{2p}^{<}(t_1,t')\\
&+\mathbf{G}_{kk'}^{<}(t,t_1)t_{2pk'}^{\dagger}(t_1)\mathbf{g}_{2p}^{A}(t_1,t')\Big],
\end{split}
\end{equation}
here, $\mathbf{G}_{kk'}^{<}(t,t')=i\langle[\tilde{d}_k^{\dagger}(t')]^T[\tilde{d}_{k'}(t)]^{T}\rangle^T$
and $\mathbf{G}_{kk'}^{r}(t,t')=-i\Theta(t-t')\langle\{\tilde{d}_k(t)\tilde{d}_{k'}^{\dagger}(t')+
[\tilde{d}_{k'}^{\dagger}(t')]^T[\tilde{d}_k(t)]^T\}\rangle^{T}$
are the lesser and retarded Green's functions of SC,
and $\mathbf{g}_{2p}^{A,<}(t,t')$ is the bare Green's function in the edge 2 which is
diagonal in Nambu space. The bare Green's functions in edge $\alpha$ are given by
\begin{equation}\label{bare time Green Fun}
\begin{split}
\mathbf{g}_{\alpha p}^{<}(t,t')=&i\langle(\tilde{c}_{\alpha p}^{\dagger})^T(\tilde{c}_{\alpha p})^T\rangle^T\\
=&\mathrm{Diag}[if_{\alpha}(\epsilon_{\alpha}^{p\sigma})e^{-i\epsilon_{\alpha}^{p\sigma}(t-t')},
if_{\alpha}(\epsilon_{\alpha}^{p\bar{\sigma}})e^{-i\epsilon_{\alpha}^{p\bar{\sigma}}(t-t')},\\
&i[1-f_{\alpha}(\epsilon_{\alpha}^{\bar{p}\sigma})]e^{i\epsilon_{\alpha}^{\bar{p}\sigma}(t-t')},
i[1-f_{\alpha}(\epsilon_{\alpha}^{\bar{p}\bar{\sigma}})]e^{i\epsilon_{\alpha}^{\bar{p}\bar{\sigma}}(t-t')}],\\
\mathbf{g}_{\alpha p}^{r,a}(t,t')=&\mp i\Theta(\pm t\mp t')\langle[\tilde{c}_{\alpha p}\tilde{c}_{\alpha p}^{\dagger}
+(\tilde{c}_{\alpha p}^{\dagger})^T(\tilde{c}_{\alpha p})^T]\rangle^T\\
=&\mp i\Theta(\pm t\mp t')\times \mathrm{Diag}[e^{-i\epsilon_{\alpha}^{p\sigma}(t-t')},\\
&e^{-i\epsilon_{\alpha}^{p\bar{\sigma}}(t-t')},e^{i\epsilon_{\alpha}^{\bar{p}\sigma}(t-t')},
e^{i\epsilon_{\alpha}^{\bar{p}\bar{\sigma}}(t-t')}],\\
\end{split}
\end{equation}
where $\epsilon_{\alpha}^{p\sigma}$ is the energy of an electron in edge $\alpha$ with momentum $p$
and spin $\bm{\sigma}$ with $\bm{\sigma}=\mathbf{m} (\mathbf{n})$ for $\alpha=1 (2)$, and
$f_{\alpha}(\epsilon)$ is the Fermi-Dirac distribution function. Substituting
Eqs.\ (\ref{analytic continuation}) and (\ref{bare time Green Fun}) into Eq.\ (\ref{cur1}), one can
derive the current
\begin{equation}\label{cur2}
\begin{split}
I_2=&\frac{ie}{\hbar}\sum_{kk'}\int \frac{d\epsilon}{2\pi}\mathrm{Tr}\Big\{\tau_z\mathbf{\Gamma}_{2}^{k'k}
\mathbf{G}_{kk'}^{<}(\epsilon)\\
&+\tau_z\mathbf{\Gamma}_{2}^{k'k}\mathbf{f}_2(\epsilon)[\mathbf{G}_{kk'}^{r}(\epsilon)-
\mathbf{G}_{kk'}^{a}(\epsilon)]\Big\},\\
\end{split}
\end{equation}
where $\mathbf{\Gamma}_{2}^{k'k}=2\pi t_{k'2p}\rho_{2}(\epsilon)t_{2pk}$ is the linewidth function, and
$\rho_2(\epsilon)=\mathrm{Diag}[\rho_{2n}^e(\epsilon),\rho_{2\bar{n}}^e(\epsilon),\rho_{2n}^h(\epsilon),
\rho_{2\bar{n}}^h(\epsilon)]$, with $\rho_{2n(\bar{n})}^{e}(\epsilon)$ and
$\rho_{2n(\bar{n})}^{h}(\epsilon)=\rho_{2n(\bar{n})}^{e}(-\epsilon)$ being the electron and hole
densities of states, respectively. In the presence of time reversal symmetry, the density of states is
the same for different spins $\rho_{2n}=\rho_{2\bar{n}}$. Therefore, the linewidth function is diagonal
$\mathbf{\Gamma}_2^{k'k}=\mathrm{Diag}[\Gamma_{2e}^{k'km},\Gamma_{2e}^{k'k\bar{m}},\Gamma_{2h}^{\bar{k'}\bar{k}m},
\Gamma_{2h}^{\bar{k'}\bar{k}\bar{m}}]$, with $\Gamma_{2e}^{k'km}=2\pi T_{k'2p}T_{2pk}[\rho_{2n}^{e}\sin^2{(\theta/2)}+\rho_{2\bar{n}}^{e}\cos^2{(\theta/2)}]$,
$\Gamma_{2e}^{k'k\bar{m}}=2\pi T_{k'2p}T_{2pk}[\rho_{2n}^{e}\cos^2{(\theta/2)}+\rho_{2\bar{n}}^{e}\sin^2{(\theta/2)}]$,
$\Gamma_{2h}^{\bar{k'}\bar{k}m}=2\pi
T_{\bar{k'}2\bar{p}}T_{2\bar{p}\bar{k}}[\rho_{2n}^{h}\sin^2{(\theta/2)}+\rho_{2\bar{n}}^{h}\cos^2{(\theta/2)}]$,
and $\Gamma_{2h}^{\bar{k'}\bar{k}\bar{m}}=2\pi T_{\bar{k'}2\bar{p}}T_{2\bar{p}\bar{k}}[\rho_{2n}^{h}\cos^2{(\theta/2)}+\rho_{2\bar{n}}^{h}\sin^2{(\theta/2)}]$.
In the following calculations, we assume that the tunneling strength is independent of $p$
and $k$, and denote the line width function as $\mathbf{\Gamma}_2^{kk'}=\Gamma_2$.
$\mathbf{f}_2(\epsilon)=\mathrm{Diag}[f_2^e(\epsilon),f_2^e(\epsilon),
f_2^h(\epsilon),f_2^h(\epsilon)]$
is the distribution function matrix with $f_{2}^{e}(\epsilon)=f_{2}(\epsilon)$ and
$f_{2}^{h}(\epsilon)=1-f_{2}(-\epsilon)$.

The lesser Green's function can be solved by the Keldysh equation
\begin{equation}\label{Keldysh 1}
\mathbf{G}_{kk'}^{<}(\epsilon)=\sum_{k_1k_2}\mathbf{G}_{kk_1}^r
\mathbf{\Sigma}_{k_1k_2}^{<}\mathbf{G}_{k_2k'}^a,
\end{equation}
where the self-energy $\mathbf{\Sigma}_{k_1k_2}^{<}(\epsilon)=\sum_{\alpha p}t_{k_1\alpha p}
\mathbf{g}_{\alpha p}^{<}(\epsilon)t_{\alpha pk_2}$ is due to the coupling with the two edges.
Using Eq.\ (\ref{bare time Green Fun}), we have
\begin{equation}\label{bare energy Green Fun}
\begin{split}
\mathbf{g}_{\alpha p}^{<}(\epsilon)=&\int dte^{i\epsilon t}\mathbf{g}_{\alpha p}^{<}(t)\\
=&2\pi i\times\mathrm{Diag}[f_{\alpha}(\epsilon_{\alpha}^{p\sigma})\delta(\epsilon-\epsilon_{\alpha}^{p\sigma}),
f_{\alpha}(\epsilon_{\alpha}^{p\bar{\sigma}})\delta(\epsilon-\epsilon_{\alpha}^{p\bar{\sigma}}),\\
&[1-f_{\alpha}(\epsilon_{\alpha}^{\bar{p}\sigma})]\delta(\epsilon+\epsilon_{\alpha}^{\bar{p}\bar{\sigma}}),
[1-f_{\alpha}(\epsilon_{\alpha}^{\bar{p}\bar{\sigma}})]\delta(\epsilon+\epsilon_{\alpha}^{\bar{p}\bar{\sigma}})],
\end{split}
\end{equation}
\begin{equation}
\begin{split}
\mathbf{g}_{\alpha p}^{r,a}(\epsilon)=&\int dte^{i\epsilon t}\mathbf{g}_{\alpha p}^{r,a}(t)\\
=&\mathrm{Diag}[(\epsilon-\epsilon_{\alpha}^{p\sigma}\pm i0^+)^{-1},
(\epsilon-\epsilon_{\alpha}^{p\bar{\sigma}}\pm i0^+)^{-1},\\
&(\epsilon+\epsilon_{\alpha}^{\bar{p}\sigma}\pm i0^+)^{-1},
(\epsilon+\epsilon_{\alpha}^{\bar{p}\bar{\sigma}}\pm i0^+)^{-1}],
\end{split}
\end{equation}
which yield self-energy
\begin{equation}
\begin{split}
\mathbf{\Sigma}_{kk'}^{<}=i\Big[\mathbf{\Gamma}_1^{kk'}\mathbf{f}_1+\mathbf{\Gamma}_2^{kk'}\mathbf{f}_2\Big],
\mathbf{\Sigma}^{r,a}_{kk'}=\mp\frac{i}{2}\Big[\mathbf{\Gamma}_1^{kk'}+\mathbf{\Gamma}_2^{kk'}\Big],
\end{split}
\end{equation}
with $\mathbf{\Gamma}_{1}^{k'k}=2\pi t_{k'1p}\rho_{1}(\epsilon)t_{1pk}$.
Applying the Dyson equation $(\mathbf{G}_{kk'}^{r,a})^{-1}=(\mathbf{g}^{r,a}_{k})^{-1}-\mathbf{\Sigma}_{kk'}^{r,a}$
with $\mathbf{g}^{r,a}_{k}$ the bare Green's function, one can obtain
\begin{equation}\label{Keldysh 2}
\mathbf{G}_{kk'}^r-\mathbf{G}_{kk'}^a=\sum_{k_1k_2}\mathbf{G}_{kk_1}^r
[\mathbf{\Sigma}_{k_1k_2}^r-\mathbf{\Sigma}_{k_1k_2}^a]\mathbf{G}_{k_2k'}^a.
\end{equation}
Inserting Eqs.\ (\ref{Keldysh 1}) and (\ref{Keldysh 2}) into Eq.\ (\ref{cur2}) yields
\begin{widetext}
\begin{equation}\label{curr4}
\begin{split}
I_2=&\frac{e}{h}\int d\epsilon\mathrm{Tr}\{
(f_2^e-f_1^e)\mathbf{\Gamma}_2^{ee}\mathbf{G}^r_{ee}\mathbf{\Gamma}_1^{ee}\mathbf{G}^a_{ee}
+(f_2^e-f_1^h)\mathbf{\Gamma}_2^{ee}\mathbf{G}^r_{eh}\mathbf{\Gamma}_1^{hh}\mathbf{G}^a_{he}
+(f_2^e-f_2^h)\mathbf{\Gamma}_2^{ee}\mathbf{G}^r_{eh}\mathbf{\Gamma}_2^{hh}\mathbf{G}^a_{he}\\
&-(f_2^h-f_1^e)\mathbf{\Gamma}_2^{hh}\mathbf{G}^r_{he}\mathbf{\Gamma}_1^{ee}\mathbf{G}^a_{eh}
-(f_2^h-f_1^h)\mathbf{\Gamma}_2^{hh}\mathbf{G}^r_{hh}\mathbf{\Gamma}_1^{hh}\mathbf{G}^a_{hh}
-(f_2^h-f_2^e)\mathbf{\Gamma}_2^{hh}\mathbf{G}^r_{he}\mathbf{\Gamma}_1^{ee}\mathbf{G}^a_{eh}\},\\
\end{split}
\end{equation}
\end{widetext}

Assuming a bias voltage $eV$ is applied to the edge 1, the edge 2 and the SC are grounded. Thus at
low temperature limit, we only need to consider the case of electrons incident from edge 1.
Since the edge states are helical, the incident state can only be the spin state $|\mathbf{m}\rangle$.
Then we can derive the coefficients of electron transmission and CAR processes
\begin{equation}\label{Coeff of elec trans}
\begin{split}
T_{\mathrm{N}}(\epsilon)=&\mathrm{Tr}[\mathbf{\Gamma}_{2}^{ee}\mathbf{G}^{r}_{ee}\mathbf{\Gamma}_{1}^{ee}
\mathbf{G}_{ee}^{a}]\\
=&\Gamma_{1}\Gamma_{2}\sum_{kk_1k_2k'}\sum_{\sigma_1\sigma_2}
[G^{ree}_{kk_1\sigma_2m}G_{k_2k'm\sigma_1}^{aee}],
\end{split}
\end{equation}
\begin{equation}\label{Coeff of CAR}
T_{\mathrm{CAR}}(\epsilon)=\Gamma_{1}\Gamma_{2}\sum_{kk_1k_2k'}\sum_{\sigma_1\sigma_2}
[G^{rhe}_{kk_1\sigma_2m}G_{k_2k'm\sigma_1}^{aeh}].
\end{equation}

In what follows we calculate the full Green's function in the SC taking
into account the coupling effect. The full Green's function in the SC is
\begin{equation}\label{self-con eq}
\mathbf{G}_{kk'}^r(\epsilon)=\mathbf{g}_k^r(\epsilon)\delta_{kk'}+\mathbf{g}_k^r(\epsilon)
\mathbf{\Sigma}^r(\epsilon)\sum_{k_1}\mathbf{G}_{k_1k'}^r(\epsilon),
\end{equation}
where the self-energy subscripts $k$ and $k'$ are omitted, and the bare Green's function $\mathbf{g}_k^r(\epsilon)$ is given by
\begin{equation}
\begin{split}
\mathbf{g}_{k}^r(\epsilon)=&[\epsilon+i0^{+}-\mathcal{H}_\mathrm{S}]^{-1}\\
=&\frac{\epsilon+i0^++\epsilon_k\tau_z-\Delta\tau_y\sigma_y}{(\epsilon+i0^+)^2-\epsilon_k^2-\Delta^2}.
\end{split}
\end{equation}
In order to solve the self-consistent Eq.\ (\ref{self-con eq}), we denote the summation as
\begin{equation}
\mathbf{\Lambda}_{k'}^r(\epsilon)=\sum_{k_1}\mathbf{G}_{k_1k'}^r(\epsilon),\nonumber
\end{equation}
then the Eq.\ (\ref{self-con eq}) reduces to
\begin{equation}\label{self-con eq 2}
\mathbf{G}_{kk'}^r(\epsilon)=\mathbf{g}_k^r(\epsilon)\delta_{kk'}+\mathbf{g}_k^r(\epsilon)
\mathbf{\Sigma}^r(\epsilon)\mathbf{\Lambda}_{k'}^r(\epsilon).
\end{equation}
Sum over $k$ both sides $\mathbf{\Lambda}_{k'}^r(\epsilon)=\mathbf{g}_{k'}^r(\epsilon)+
\Big[\sum_k\mathbf{g}_k^r(\epsilon)\Big]\mathbf{\Sigma}^r(\epsilon)\mathbf{\Lambda}_{k'}^r(\epsilon)$,
we can obtain
\begin{equation}
\mathbf{\Lambda}_{k'}^r(\epsilon)=\frac{1}{1-\Big[\sum_k\mathbf{g}_k^r(\epsilon)\Big]
\mathbf{\Sigma}^r(\epsilon)}\mathbf{g}_{k'}^r(\epsilon).
\end{equation}
By summing up both sides of Eq.\ (\ref{self-con eq 2}) over $k$ and $k'$, one can obtain
\begin{equation}\label{Sum of ful Green}
\begin{split}
\sum_{kk'}\mathbf{G}_{kk'}^r(\epsilon)=&\sum_k\mathbf{g}_k^r(\epsilon)+\Big[\sum_k\mathbf{g}_k^r(\epsilon)
\Big]\mathbf{\Sigma}^r(\epsilon)\\
&\times\frac{1}{1-\Big[\sum_k\mathbf{g}_k^r(\epsilon)\Big]\mathbf{\Sigma}^r(\epsilon)}\Big[\sum_{k'}
\mathbf{g}_{k'}^r(\epsilon)\Big].
\end{split}
\end{equation}
The sum of the bare Green's function $\mathbf{g}_{k}^r(\epsilon)$ over $k$ is
\begin{equation}\label{Sum of Green}
\begin{split}
\sum_k\mathbf{g}_{k}^r(\epsilon)=&\int d\epsilon_k\rho_{\mathrm{S}}(\epsilon_k)\mathbf{g}_k^r(\epsilon)\\
=&-i\pi\rho_{\mathrm{S}}^0\frac{\epsilon-\Delta\tau_y\sigma_y}{\sqrt{\epsilon^2-\Delta^2}}
=-i\pi\rho_{\mathrm{S}}(1-\frac{\Delta}{\epsilon}\tau_y\sigma_y).
\end{split}
\end{equation}
where $\rho_{\mathrm{S}}$ is the density of the states of the quasi-particle in the SC, and
$\rho_{\mathrm{S}}^0$ is the normal metallic state density.
Substituting Eq.\ (\ref{Sum of Green}) into Eq.\ (\ref{Sum of ful Green}) yields
\begin{equation}\label{Sum of ful Green 2}
\begin{split}
\sum_{kk'}\mathbf{G}_{kk'}^r(\epsilon)=-\frac{i\pi\rho_{\mathrm{S}}[1+\frac{\pi\rho_{\mathrm{S}}\Gamma}{2}
(1-\frac{\Delta^2}{\epsilon^2})]}{\Upsilon}
+\frac{i\pi\rho_{\mathrm{S}}\frac{\Delta}{\epsilon}}{\Upsilon}\tau_y\sigma_y,
\end{split}
\end{equation}
where $\Upsilon=1+\pi\rho_{\mathrm{S}}\Gamma+\frac{\pi^2\rho_{\mathrm{S}}^2\Gamma^2}{4}
(1-\frac{\Delta^2}{\epsilon^2})$ and $\Gamma=\Gamma_1+\Gamma_2$.
By inserting Eq.\ (\ref{Sum of ful Green 2}) into the Eqs.\ (\ref{Coeff of elec trans}) and (\ref{Coeff of CAR}),
one can obtain the tunneling coefficients
\begin{equation}\label{CNR}
T_{\mathrm{N}}(\epsilon)=\cos^2\frac{\theta}{2}\Gamma_1\Gamma_2\Big|-\frac{i\pi\rho_{\mathrm{S}}
[1+\frac{\pi\rho_{\mathrm{S}}\Gamma}{2}
(1-\frac{\Delta^2}{\epsilon^2})]}{\Upsilon}\Big|^2,
\end{equation}
\begin{equation}\label{CAR}
T_{\mathrm{CAR}}(\epsilon)=\sin^2\frac{\theta}{2}\Gamma_1\Gamma_2\Big|\frac{i\pi\rho_{\mathrm{S}}
\frac{\Delta}{\epsilon}}{\Upsilon}\Big|^2.
\end{equation}
The two scattering probabilities incorporate the relative spin rotation angle $\theta$ induced by the Rashba
SOC in the helical edge states. This indicates that the manipulation of the edge state
spin can be inferred from these probabilities, representing the main physical result of this paper.

\section{Results and Discussions}
The electron transmission probability (\ref{CNR}) and CAR probability (\ref{CAR}) between the two sets
of edge states provide a direct measure of our control over the helical edge states, as illustrated in
Fig.\ \ref{Fig1}. We first examine the extreme cases where the Rashba SOC strengths
$\lambda_{1,2}$ are 0 and infinity, corresponding to rotation angles $\theta_{1,2}=0$ and
$\theta_{1,2}=\pi/2$, respectively. In the absence of Rashba SOC, the probability of CAR
is zero, and only electron transmission occurs at edge 2. This is because the incident electrons at edge
1 have spin-up polarization, while the outgoing states at edge 2 also possess spin-up polarization,
resulting in a lack of spin-paired electrons to match the incident electrons and thereby preventing CAR.
When Rashba SOC is present, it causes the spins of the two edges to rotate in opposite directions.
As the Rashba SOC strength approaches infinity, the spin of the incident electron at edge
1 rotates clockwise by $\pi/2$ from spin-up to spin-right, while the spin of the outgoing state at edge 2
rotates counterclockwise from spin-up to spin-left, precisely opposite to the spin of the incident electron.
This alignment enables the formation of Cooper pairs and facilitates CAR process. At this extreme, the
outgoing spin is antiparallel to the incident spin, rendering the two states orthogonal and thus prohibiting
electron transmission.
\begin{figure}[b!]
\begin{center}
\includegraphics[width=1\columnwidth]{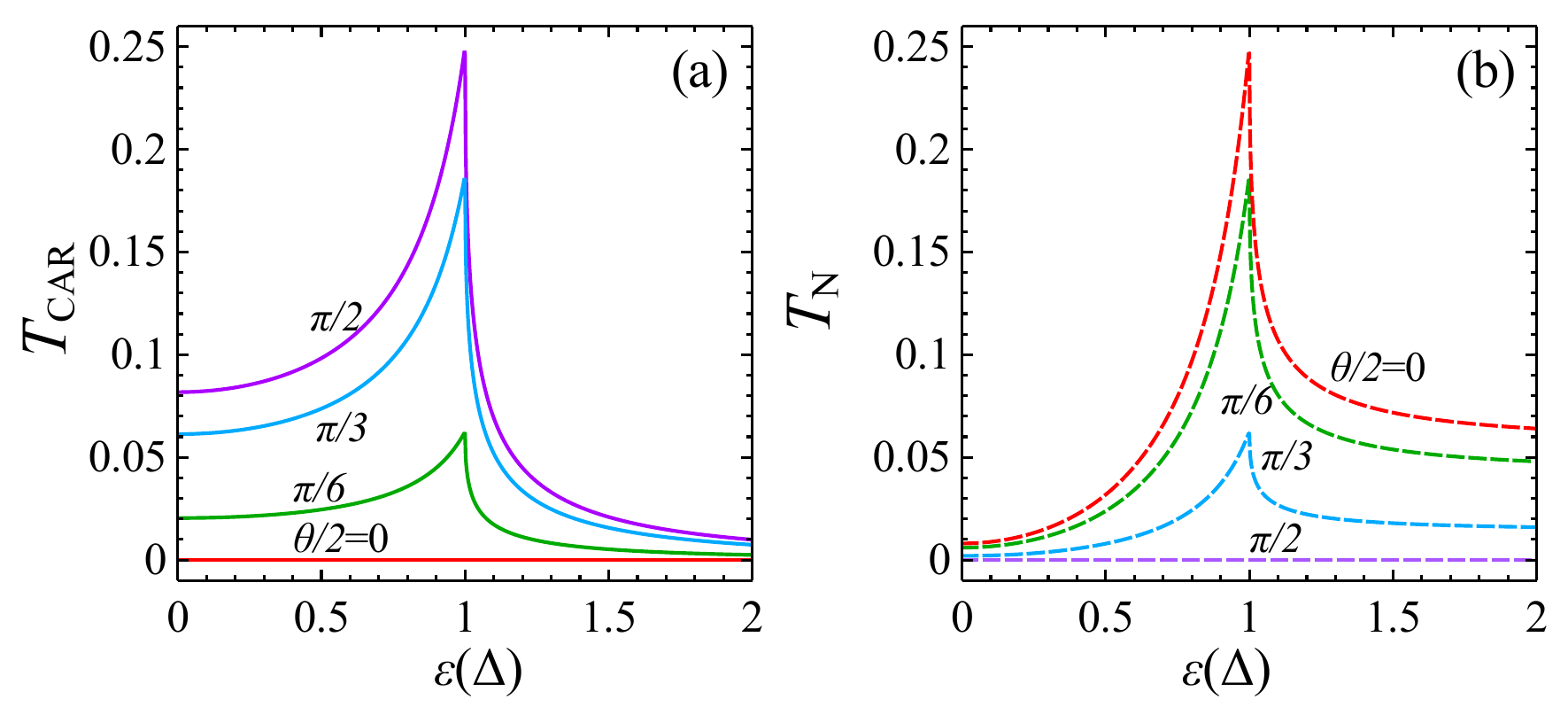}\\[0pt]
\caption{(Color online) The CAR probability (a) and normal electron transmission probability (b) as functions of
incident energy for different rotation angles $\theta$. The linewidth functions are set to $\Gamma_1=\Gamma_2=0.1$.}
\label{Fig2}
\end{center}
\end{figure}

In general cases, Rashba SOC induces a clockwise rotation of the electron spin on edge 1 by an
angle $\theta_1$, aligning the spin of the incident electron parallel to the direction $\mathbf{m}$.
Conversely, the spin of the outgoing electron is oriented parallel to $-\mathbf{m}$. On edge 2,
Rashba SOC causes the electron spin to rotate counterclockwise by an angle $\theta_2$, with
the spin of the reflected electron parallel to the direction $-\mathbf{n}$. Thus, the electron
transmission amplitude on edge 2 is proportional to $\langle \mathbf{m}|-\mathbf{n}\rangle=\cos{\frac{\theta}{2}}$.
CAR necessitates the formation of Cooper pairs with spins oriented oppositely, so the outgoing hole's spin
aligns with $\mathbf{m}$ (corresponding to the vacancy of electron with spin state $|-\mathbf{m}\rangle$).
The amplitude for this process is proportional to
$\langle -\mathbf{m}|-\mathbf{n}\rangle=\langle \mathbf{m}|\mathbf{n}\rangle=\sin{\frac{\theta}{2}}$. Consequently, CAR provides a direct measure of the
relative spin deflection angle $\theta=\theta_1+\theta_2$ between the edge state spins at edges 1 and 2 under
the influence of the Rashba SOC. Fig.\ \ref{Fig2} illustrates the variation of the CAR probability
$T_{\mathrm{CAR}}(\epsilon)$ and the electron transmission probability $T_{\mathrm{N}}(\epsilon)$ as functions
of incident energy for different angles $\theta$. It is observed that both probabilities reach their maximum as
the incident energy approaches the superconducting gap $\Delta$, because the wave function decays more slowly in the SC
when the energy is near the superconducting gap. Additionally, the CAR probability increases from zero to a finite
value with increasing $\theta$, while the probability of electron transmission decreases. These trends
reflect the influence of Rashba SOC on the spin deflection of edge states.

\begin{figure}[t!]
\begin{center}
\includegraphics[width=0.77\columnwidth]{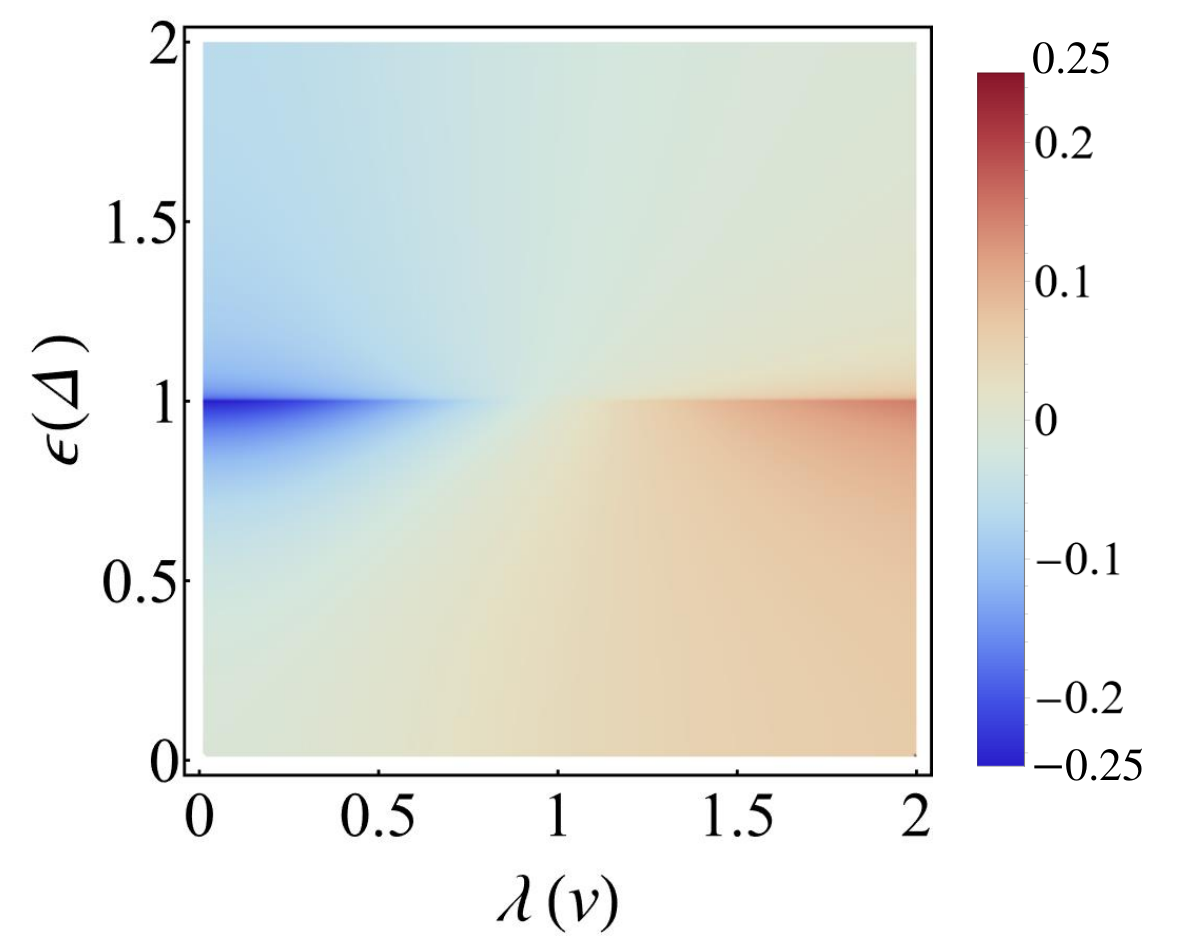}\\[0pt]
\caption{(Color online) Differential conductance $G_2$ as functions of incident energy and Rashba SOC strength.
The linewidth functions are set to $\Gamma_1=\Gamma_2=0.1$.}
\label{Fig3}
\end{center}
\end{figure}

The conductance of edge 2 is given by
\begin{equation}\label{conduc}
G_2=\frac{e^2}{h}(T_{\mathrm{CAR}}-T_{\mathrm{N}}).
\end{equation}
Fig.\ \ref{Fig3} illustrates the dependence of conductance on the strength of Rashba SOC and
incident energy. Notably, when $\lambda/v > 1$, a positive conductance is observed if the incident energy falls
within the superconducting gap $\epsilon<\Delta$, indicating the dominance of CAR. When $\lambda/v<1$,
positive conductance appears only within low energy regions. Regardless of whether the conductance is positive, the control of the edge state spin will always be reflected through the change in conductance as the Rashba SOC varies.

Specifically, when the incident energy $\epsilon\rightarrow\Delta$, the transmission coefficient (\ref{CNR})
and CAR coefficient (\ref{CAR}) simplify to $T_N=\cos^2(\theta/2)\Gamma_1\Gamma_2/\Gamma^2$
and $T_A=\sin^2(\theta/2)\Gamma_1\Gamma_2/\Gamma^2$, respectively. Consequently, the conductance reduces to
\begin{equation}\label{conduc1}
G_2=-\frac{e^2}{\hbar}\frac{\cos{\theta}\Gamma_1\Gamma_2}{\Gamma^2}.
\end{equation}
The conductance is directly related to the cosine of the relative rotation angle between
the spins at the two edges, thereby clearly reflecting the influence of Rashba SOC on the
modulation of edge state spin. The theoretical model encompassing helical edge states coupled with an
$s$-wave SC heterostructure, the modulation of Rashba SOC through gate voltage,
and the measurement of multi-terminal electron transport are all achievable with current experimental
techniques. Thus, the theoretical scheme and results we present are directly applicable and realizable
within existing experimental setups.

\section{Summary}
In this paper, we propose utilizing CAR to assess the efficiency of gate voltage control on edge state spin.
We calculate the transport properties of a QSHIs-SC heterojunction using non-equilibrium
Green's function method. Our results indicate that the CAR probability, electron transmission probability, and
non-local conductance all incorporate the relative spin rotation angle $\theta$, which arises from Rashba
SOC in the helical edge states. Specifically, as the incident energy approaches the
superconducting gap $\epsilon\rightarrow\Delta$, the differential conductance $G_2\propto-\cos{\theta}$. Consequently,
the influence of the Rashba SOC on edge state spin can be quantitatively evaluated through CAR, offering
a practical approach for experimental validation.

\begin{acknowledgments}
This work was supported by the National Natural Science Foundation of China under Grant No. 12222406 (W.C.), No. 12074172 (W.C.),
and No. 12264019 (W.L.), the Natural Science Foundation of Jiangsu Province under Grant No. BK20233001 (W.C.),
the Fundamental Research Funds for the Central Universities under No. 2024300415 (W.C.), the National Key Projects for Research and Development of China under No. 2022YFA120470 (W.C.), and the Jiangxi provincial Department of Science and Technology under Grant No. 20242BAB23005 (W.L.).
\end{acknowledgments}

\end{document}